\documentclass{elsart}

\usepackage{amssymb}
\usepackage{graphicx}
\usepackage{subfigure}
\usepackage{amsmath}
\begin{document}

\begin{frontmatter}



\title{On-site underground background measurements for the KASKA
reactor-neutrino experiment}


\author[TIT]{H.~Furuta},
\author[TMU]{K.~Sakuma\corauthref{cor1}},
\author[NGT]{M.~Aoki},
\author[MYK]{Y.~Fukuda},
\author[TIT]{Y.~Funaki},
\author[KOB]{T.~Hara},
\author[TMU]{T.~Haruna},
\author[KEK]{N.~Ishihara},
\author[NGT]{M.~Katsumata},
\author[NGT]{T.~Kawasaki},
\author[TIT]{M.~Kuze},
\author[TIT]{J.~Maeda},
\author[TIT]{T.~Matsubara},
\author[TMU]{T.~Matsumoto\corauthref{cor2}},
\author[NGT]{H.~Miyata},
\author[HIT]{Y.~Nagasaka},
\author[TMU]{T.~Nakagawa\corauthref{cor3}},
\author[NGT]{N.~Nakajima},
\author[TIT]{K.~Nitta},
\author[NGT]{K.~Sakai},
\author[TGK]{Y.~Sakamoto},
\author[THK]{F.~Suekane},
\author[TMU]{T.~Sumiyoshi},
\author[THK]{H.~Tabata},
\author[NGT]{N.~Tamura},
\author[THK]{Y.~Tsuchiya}\\

\address[TIT]{Department of Physics, Tokyo Institute of Technology, Tokyo 152-8551 Japan}
\address[TMU]{Department of Physics, Tokyo Metropolitan University, Hachioji 192-0397 Japan}
\address[NGT]{Department of Physics, Niigata University, Niigata 950-2181 Japan}
\address[MYK]{Department of Physics, Miyagi University of Education, Sendai 980-0845 Japan}
\address[KOB]{Department of Physics, Kobe University, Kobe 657-8501 Japan}
\address[KEK]{Institute of Particle and Nuclear Studies, High Energy
 Accelerator Research Organization (KEK), Tsukuba 305-0801 Japan}
\address[HIT]{Department of Computer Science, Hiroshima Institute of Technology, Hiroshima 731-5193 Japan}
\address[TGK]{Department of Information Science, Tohoku Gakuin University, Sendai 981-3193 Japan}
\address[THK]{Research Center for Neutrino Science, Tohoku University, Sendai 980-8578 Japan}
\corauth[cor1]{Present address: Ministry of Economy, Trade and Industry
 of Japan, Japan.}
\corauth[cor2]{~Present address: Institute of Particle and Nuclear
 Studies, KEK, Japan.}
\corauth[cor3]{~~~Present address: Hewlett-Packard Japan, Ltd., Japan.}

\address{}

\begin{abstract}
On-site underground background measurements were performed for
the planned reactor-neutrino oscillation experiment KASKA at
Kashiwazaki-Kariwa nuclear power station in Niigata, Japan.
A small-diameter boring hole was excavated down to 70m underground
level, and a detector unit for $\gamma$-ray and cosmic-muon measurements
was placed at various depths to take data.
The data were analyzed to obtain abundance of natural radioactive
elements in the surrounding soil and rates of cosmic muons that
penetrate the overburden.
The results will be reflected in the design of the KASKA experiment.
\end{abstract}

\end{frontmatter}

\section{Introduction}
\label{sec:intro}
A new neutrino-oscillation experiment KASKA\cite{LoI} is planned at
Kashiwazaki-Kariwa nuclear power station in Niigata, Japan.
It will make precise measurements of neutrino flux from the
reactors in order to measure the yet undetected $\theta_{13}$
neutrino mixing angle. The detectors will be placed in deep
underground vertical shaft holes.  Two near detectors are planned
inside the reactor site at a depth of 50~m in separate shafts, and two far
detectors are planned outside the site, at a distance of about 1.6~km
and a depth of 150~m in one shaft hole.

\begin{figure}[htbp]
\begin{center}
\includegraphics[scale=0.5]{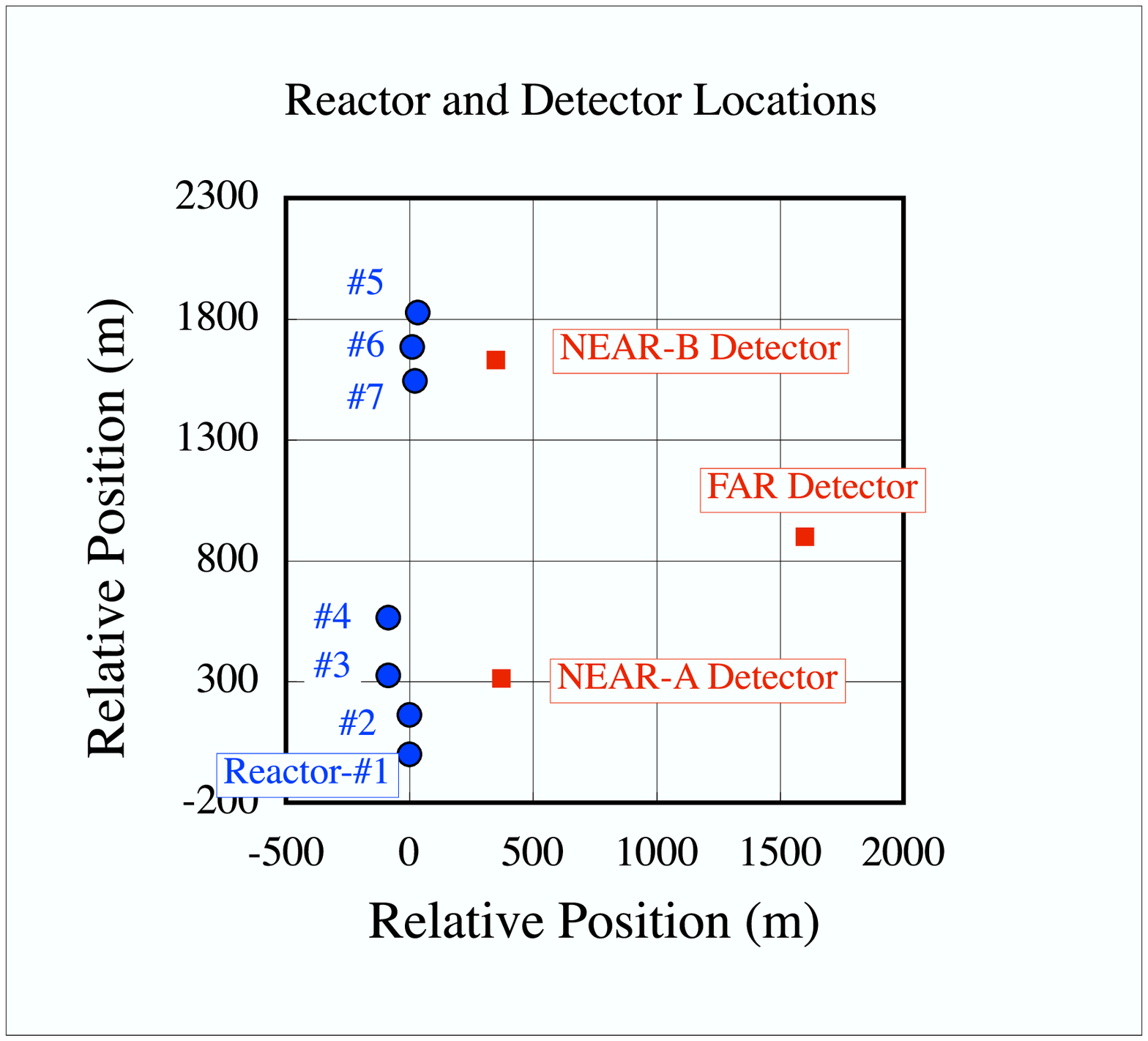}
\caption{The locations of reactors and detectors.}
\label{fig:det_pos}
\end{center}
\end{figure}

During October and November 2004, a boring study was conducted
on the reactor site, at a place where one of the near detectors
will be placed (NEAR-B position in Fig.~\ref{fig:det_pos}).
The purpose of the boring study was to obtain
the soil samples needed for the planning of civil construction
of the vertical shaft holes, and to measure backgrounds at the
actual detector location in order to reflect in the detector design.

The boring hole extended down to 70~m underground, with a diameter
of 66~mm.  After the sample cores had been taken, a detector for
measuring $\gamma$-ray and cosmic-muon background was deployed
into the hole, and data were taken at various depths.
The experimental setup of the measurement and data analyses are
described in the following sections.

\section{Experimental setup}
\label{sec:exp}
\begin{figure}[htbp]
\begin{center}
\includegraphics[scale=0.5]{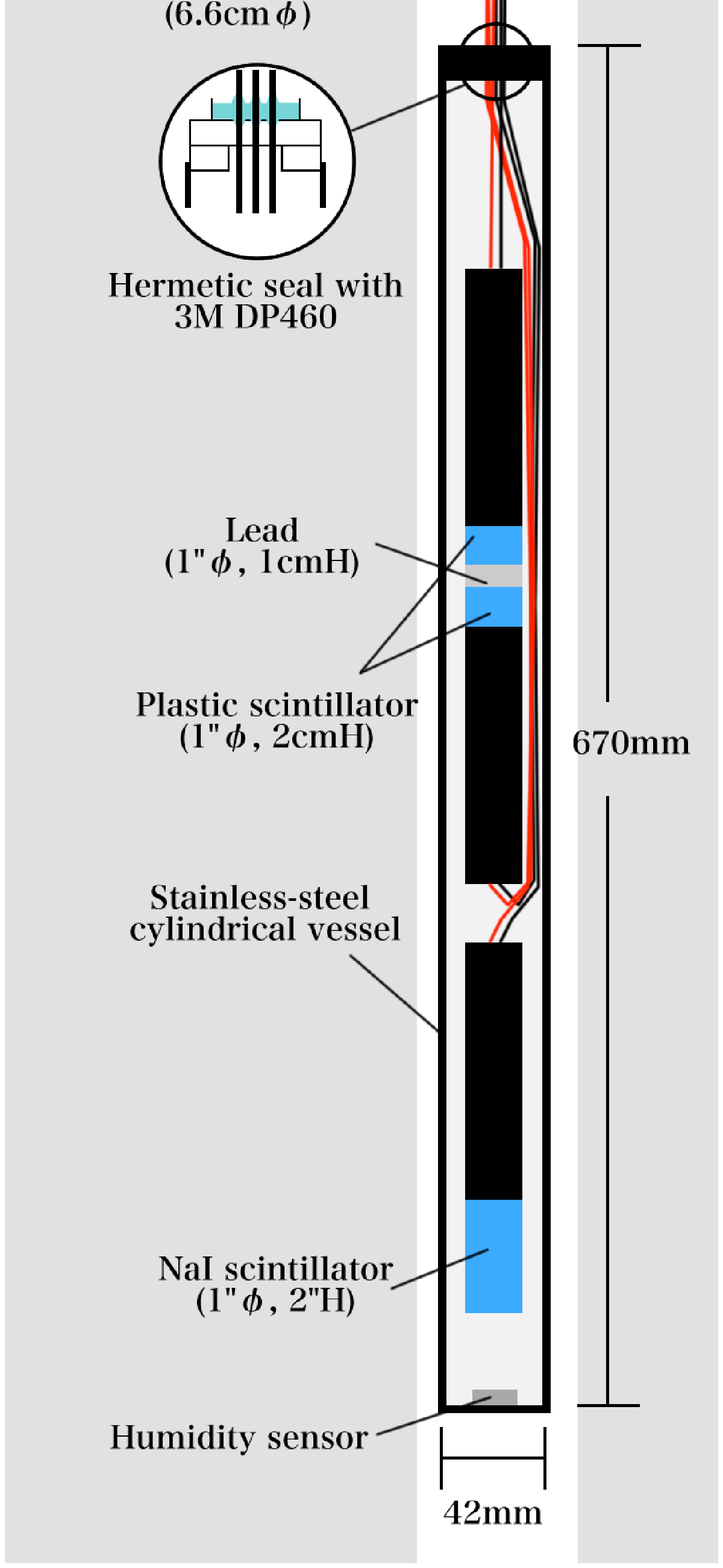}
\caption{The schematic of the detector.}
\label{fig:det}
\end{center}
\end{figure}

The detector consisted of a NaI(Tl) scintillator for the $\gamma$-ray
measurement and a pair of plastic-scintillators for the cosmic-muon
measurement.
The schematic of the detector is shown in Fig.~\ref{fig:det}.
\\

The NaI scintillator had a diameter of 1~inch and a length of 2~inches,
and was viewed by a Hamamatsu H8643 photomultiplier (PMT).
The cosmic-muon counter had two plastic scintillators of 1-inch
diameter and 2~cm thickness, sandwiching a 1~cm-thick lead plate
to reduce correlated Compton-$\gamma$ background,
and both scintillators were viewed by Hamamatsu H8643 PMTs.

The scintillators and PMTs were contained in a stainless-steel
cylindrical vessel of 42~mm diameter and 670~mm length.
Cables for high voltage(RG174/U) and signals(1.5D-QEV), 100~m long,
were taken out of the vessel through small holes, and the vessel was
made hermetic with 3M DP460 epoxy adhesive.
Also a humidity sensor was put in the vessel to detect a possible water leak.
The hermiticity of the vessel was tested up to 10 bars of water pressure.

After the core samples
were taken, PVC pipes (inner/outer diameters 51/60~mm) were put into
the hole to protect the hole surface. 
Then the detector was deployed into the hole
using a stainless wire, and data were taken at various depths,
down to 65~m from the surface.  The level of the surface was found
to be 30~m above the sea level, and the water level in the hole
was found to be approximately at the sea level.

The data were taken with either of the three PMTs triggering above a
threshold.  Pulse heights of PMTs were recorded with a VME
ADC (LeCroy 1182).  In addition, discriminator output of each PMT was
recorded with a bit register.  Single rates of cosmic-muon
PMTs were about 1~Hz, and their coincidence rate was between 4-17 events/hour,
depending on the depth.  The single rate of $\gamma$-ray PMT
was between 15-35~Hz, depending on the depth.

\section{$\gamma$-ray analysis}
\begin{figure}[htbp]
\begin{center}
\includegraphics[scale=0.65,angle=0]{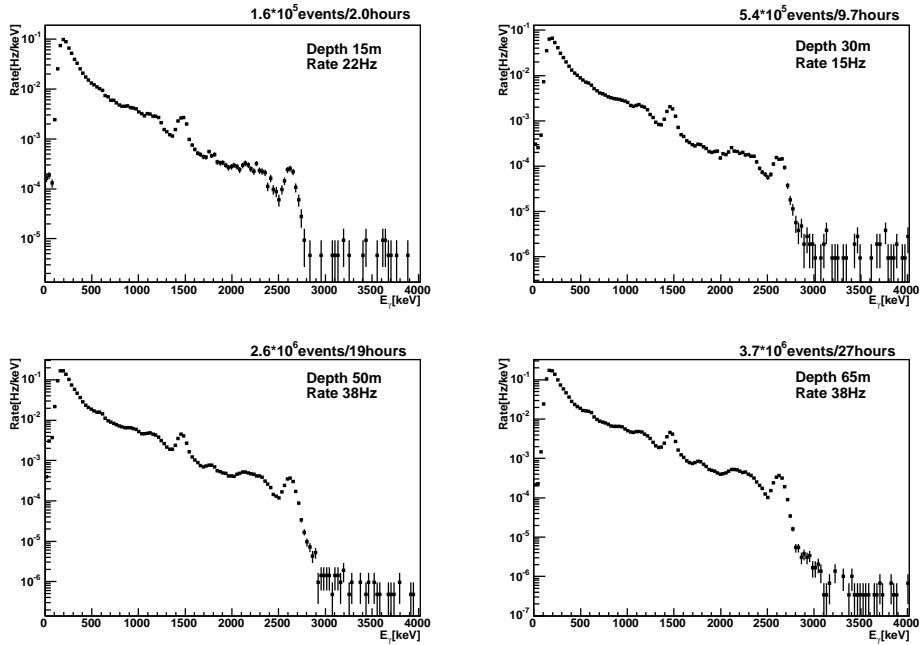}
\caption{The measured $\gamma$-ray energy
spectra at 15, 30, 50 and 65~m depths.}
\label{fig:gamma-raw}
\end{center}
\end{figure}

Figure~\ref{fig:gamma-raw} shows the measured $\gamma$-ray energy
spectra at 15, 30, 50 and 65~m depths.

Distinct peaks at 1460 and 2600~keV are seen, with other small peaks.
They correspond to 1461~keV $\gamma$'s from $^{40}\rm K$ and 2615~keV $\gamma$'s
from $^{208}\rm Tl$ in the $^{232}\rm Th$ chain.
The analysis is made under the hypothesis that the sources are
$^{238}\rm U$, $^{232}\rm Th$, and $^{40}\rm K$ contained in the soil,
and that the decay chains of $^{238}\rm U$ and $^{232}\rm Th$ series
are in radio-equilibrium.
To reproduce the spectra, Geant4~\cite{geant4} simulations are
performed, with a geometry consisting of (from outer to inner)
soil space (100~cm diameter and 100~cm height, 30\% water
concentration), PVC pipe, water, stainless-steel vessel and NaI
scintillator.
For each of $^{238}\rm U$, $^{232}\rm Th$ series, 30 $\gamma$-ray energies
having largest branching ratios are considered~\cite{energyrate},
and for $^{40}\rm K$, single
$\gamma$-ray of 1461~keV is considered in the simulation.

\begin{figure}[htbp]
\begin{center}
\includegraphics[scale=0.65,angle=0]{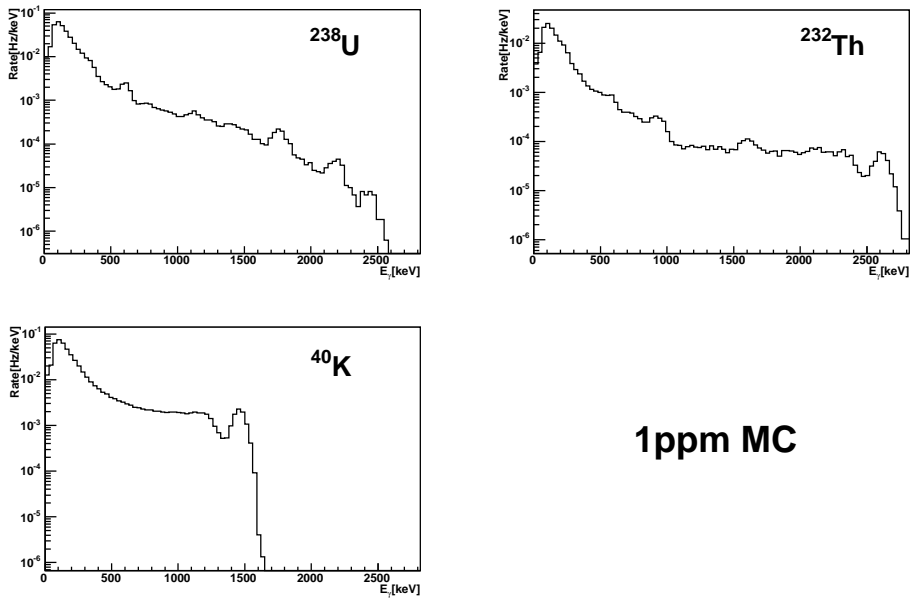}
\caption{The simulated spectra from each source
for 1ppm weight concentration in the soil.}
\label{fig:1ppm}
\end{center}
\end{figure}

The energies are smeared with a detector resolution of
$\sigma(E)/E=3.6\%/\sqrt{E(MeV)}$, which reproduces the peak at 2615~keV.
Figure~\ref{fig:1ppm} shows the simulated spectra from each source
for 1~ppm weight concentration in the soil.

\begin{figure}[htbp]
\begin{center}
\includegraphics[scale=0.65,angle=0]{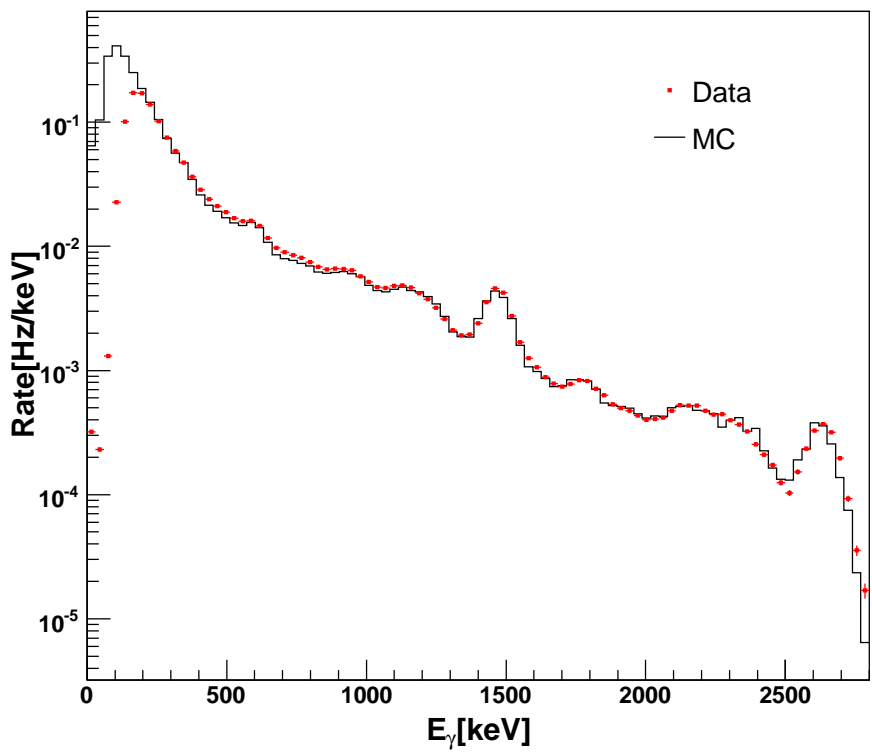}
\caption{The comparison of the data and fitted spectrum from the simulation for the depth of 65~m.}
\label{65m}
\end{center}
\end{figure}

Then, a linear combination of the three sources are fitted to the
measured spectra, and a $\chi^2$ fitting for the region above 1200~keV
is performed to obtain the weight concentrations of the three sources.
Table~\ref{tab:ppm} shows the fitting results, and Fig.~\ref{65m} shows
the comparison of the data and the fitted spectrum from the simulation for
the depth of 65~m.  A good agreement is observed.

The notable difference in $\gamma$-ray rates for depths up to 30~m and
for depths of ~50m and above is coming presumably from different
stratum structure at the location.  The weight concentrations at 50~m
are not very different from the estimation, so that it ensures the
appropriateness of the current $\gamma$-ray shield design.

\begin{table}[htpb]
\renewcommand{\arraystretch}{0.98}
\caption{Each series weight concentrations (ppm) obtained from the fitting}\label{tab:ppm}
\begin{center}
\begin{tabular}{|c|c|c|c|}\hline
Depth&$^{238}$U&$^{232}$Th&$^{40}$K\\\hline
15m&1.3$\pm$0.06&2.7$\pm$0.07&1.3$\pm$0.03\\\hline
30m&0.8$\pm$0.02&2.0$\pm$0.03&1.0$\pm$0.01\\\hline
50m&2.1$\pm$0.03&6.3$\pm$0.04&1.5$\pm$0.01\\\hline
65m&2.3$\pm$0.02&6.1$\pm$0.03&1.5$\pm$0.007\\\hline
\end{tabular}
\end{center}
\end{table}

\section{Cosmic-muon analysis}

\begin{figure}[htbp]
\begin{center}
{\includegraphics[width=.9\hsize]{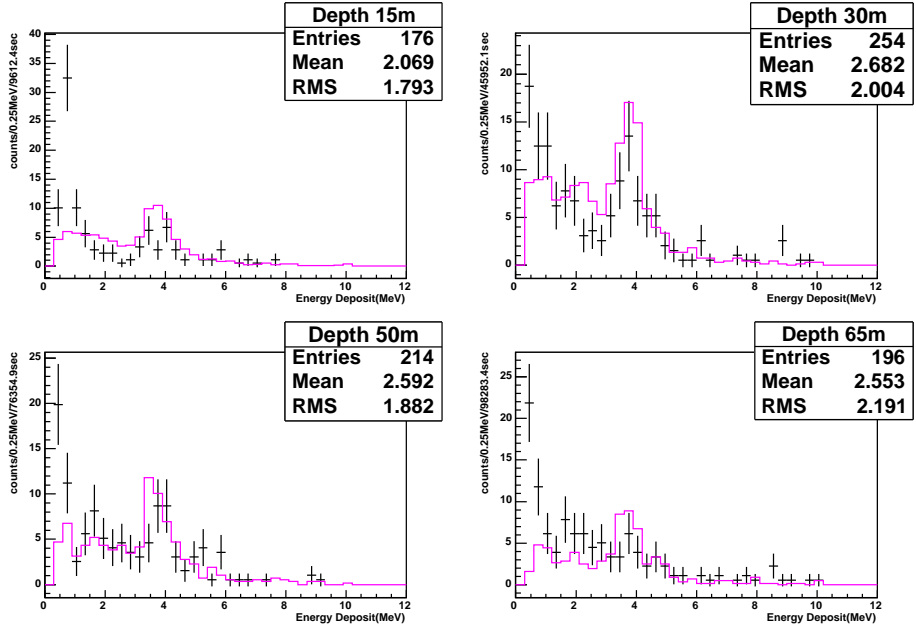}}
\caption{Energy deposit in plastic scintillators of coincident hits
 (E$_{th}$=0.4MeV).  The crosses show the measured data, while
the histograms show the Geant4 simulation.}
\label{fig:mip_c}
\end{center}
\end{figure}

\begin{figure}[htbp]
\begin{center}
 \includegraphics[width=.9\hsize]{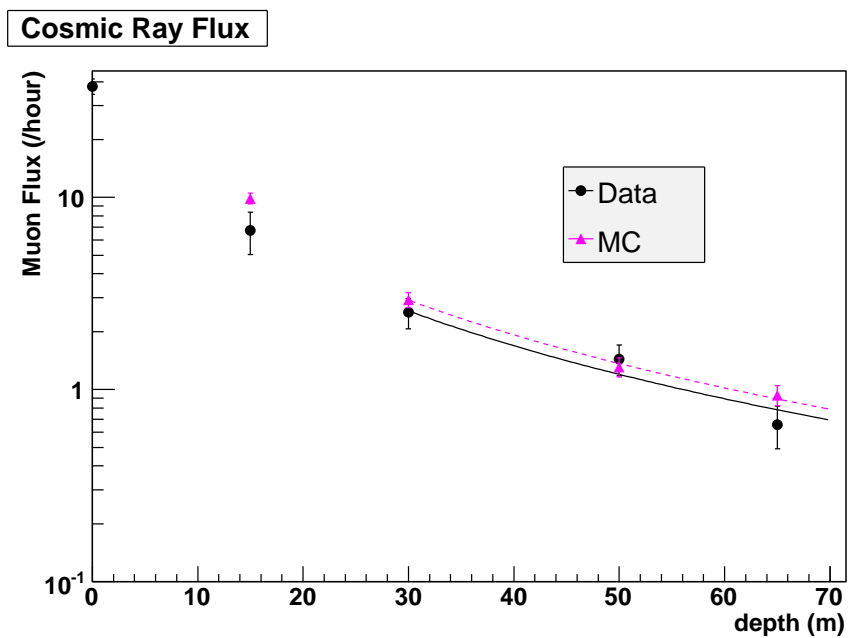}
 \caption{Measured (circles) and simulated (triangles) rates above
 3~MeV.  Solid lines are fitted functions excluding the rates at the
 surface and 15m depth.}
 \label{fig:murate}
\end{center}
\end{figure}

Figure~\ref{fig:mip_c} shows the energy deposit in the plastic scintillators
measured at 15, 30, 50 and 65~m depths,
when the coincidence of the two scintillators are taken.
A peak corresponding to minimum ionizing particles (MIP)
is clearly seen around 4~MeV, and also spectra of the
environmental $\gamma$-rays are seen at low-energy region.
The low-energy region also contains events in which the cosmic
ray traversed only a part of the scintillator.
Though the coincidence are taken, spectra of the Compton scattered
$\gamma$-rays remain.
A Geant4 simulation is compared with the measured data.
The simulation generates muon flux at the surface, with a flat
zenith-angle dependence.
The energy spectrum of the muon is parameterized as in the following equation
with the measured spectrum data of BESS~\cite{BESS} and CAPRICE~\cite{CAPRICE}
experiments. \begin{equation}\label{muFit2}
 \displaystyle \frac{dj_{\theta=0}(E) }{dE} 
 = 0.0049 \left\{ 1.057 + 0.217 E \right\}^{-3.26}\hspace{5mm} \rm (/cm^2/sr/sec/(GeV/c))
\end{equation}
Then, the muons are tracked down in the soil space beneath the surface,
and energy deposits in the scintillators are calculated.
In Fig.~\ref{fig:mip_c}, it is seen that the simulation fairly well
reproduces the energy spectrum above 3~MeV,
where the effect of environmental $\gamma $-rays is negligible.
From the comparison of the measured and simulated rates above 3~MeV, 
the muon flux at each depth is well demonstrated by the empirical formula,
which is shown in Fig.~\ref{fig:murate}. 
In the figure, fitted functions using an empirical formula~\cite{rika},
\begin{equation*}\label{muFit}
 \displaystyle I_v = \frac{1740000}{ h + 400 } \left( h + 11 \right)^{-1.53}
 \times {\rm exp}(-7.0\times 10^{-4} h) \rm (/m^2/sr/sec)
\end{equation*}
(where $h$ is the depth in mwe) are also given.
In the fitting, the data rates near the surface are
excluded since they may contain additional rates from electromagnetic showers.
The agreement between the prediction is within 20\%, and 
it ensures the estimation of the cosmic-origin background rates
in the main detector.

\section{Summary}
On-site underground background measurements were performed for
the planned reactor-neutrino oscillation experiment KASKA at
Kashiwazaki-Kariwa nuclear power station in Niigata, Japan.
Gamma-rays and cosmic-muons were measured using a small detector
deployed at various depths.  Spectra of $\gamma$-rays were fitted
and weight concentrations of radioactive elements at each depth of
the surrounding soil were measured.  Cosmic muon rates were obtained from
the minimum-ionizing peaks of the data and compared with the simulation
to evaluate its reliability.
The results will be reflected in the design of the KASKA detector.

\section*{Acknowledgements}
This work was performed under a grant-in-aid for scientific research
(\#16204015) of Ministry of Education, Culture, Sport and Technology
of Japan (MEXT).
We would like to thank Super-Kamiokande collaboration for the help
in the test of the hermeticity of the detector vessel.


\end{document}